\documentclass[12pt]{article}
\usepackage{xcolor}
\usepackage{graphicx}
\usepackage{amsmath}
\usepackage{amssymb}
\usepackage{geometry}
\usepackage{array, tabularx}
\usepackage{multirow}
\usepackage{pdflscape}
\usepackage{amsfonts}
\usepackage{placeins}
\usepackage{hyperref}
\usepackage{url}
\usepackage{caption}
\usepackage{epstopdf}
\usepackage[normalem]{ulem}
\usepackage{subcaption}
\begin{document}
\title{\large\bfseries Quantum Symmetric Cooperative Game with a Harmonious Coalition}
\date{}
\author{A.S. Elgazzar$^1$ and H.A. Elrayes$^2$}
\maketitle
\begin{center}
$^1$Mathematics Department, Faculty of Science, Arish University, 45516 Arish, Egypt\\
\url{elgazzar1@yahoo.com} \\
$^2$Mathematics Department, Faculty of Science, Damietta University, New Damietta, Egypt\\
\url{heba_elrayes@du.edu.eg}
\end{center}
\baselineskip24pt
\vspace{1cm}

\begin{abstract}
   The concept of forming harmonious coalitions is introduced to both classical and quantum symmetric cooperative game. In both cases, players are motivated to form coalitions. Also, the main feature of the cooperative game is conserved.
   \end{abstract}

\section{Introduction}
Game theory \cite{1,2} is the study of mathematical models of conflict and cooperation between rational individuals. It had been formalised by Neumann and Morgenstern \cite{1}. Game theory is applied to many areas especially social and biological sciences \cite{2}.

Cooperative game \cite{2} is an interesting element of game theory. Players can form coalitions, binding agreements, etc. Therefore, it is important to investigate forming coalitions in a cooperative game. The main feature of a cooperative game is the characteristic function of a coalition. It is the payoff which a coalition assures for itself for all strategies of the opponent. Classical cooperative game is a zero-sum game, and players are found to be motivated to form a coalition.

There is an increasing interest in applying some concepts of quantum mechanics to many scientific fields. For example, in quantum information theory \cite{3}, quantum mechanics offers new possibilities for the communication and processing information. Superposition principle and entanglement are fundamental concepts in quantum information theory.

A combination of game theory and concepts of quantum information theory leads to quantum game theory \cite{4,5}. Quantisation of games allows to expand the strategy space to a Hilbert space of strategies instead of a set of few pure classical strategies. Also, linear superpositions between strategies are allowed. The Marinatto-Weber's scheme \cite{5} is simple and straight forward. It is applied to quantise many classical games \cite{6,7,8}.

Iqbal and Toor \cite{6} had studied a quantum symmetric cooperative game. In their scheme, players are not motivated to form a coalition, and the zero-sum property is lost.

In many situations, individuals are allowed to have binding agreements and forming a coalition beforehand. This is obvious in sport competitions and political party formation. Therefore, we modify the Iqbal and Toor's scheme \cite{6} by assuming coalitions beforehand. Also, the concept of harmonious coalition is introduced. Harmonious coalition means players in the same coalition use the same strategy (pure or a superposition of strategies). Marinatto-Weber's scheme \cite{5} is used. Players are found to be motivated to form a coalition, and the zero-sum property of the game is conserved.

\section{Classical Form}
A classical three-player and zero-sum cooperative game \cite{6} with pure strategies is given as:
\begin{equation}\label{1}
  \Gamma=\langle N=\{1,2,3\} ,U_{N}=\{F,E\} ,P_{N}:U_{1} \times U_{2} \times U_{3} \longrightarrow \mathbb{R} \rangle,
\end{equation}
where $N$ is the number of players, $U$ is the strategy and $P$ is the payoff. Let $Q=\{A,B,C\}$ be the set of players. Each player chooses one strategy $F $ or $E$. If the three players choose the same strategy, then every player gets zero of money. The two players who choose the same strategy get one unit of money each from the single player. Then, the payoff matrix is given as follows:
\FloatBarrier
\begin{table}[ht]
\begin{center}
\caption{Payoff matrix for the cooperative game, when player C plays strategy F.}\label{2}
\begin{tabular}{lll}
\hline
& \quad \quad\quad player B: F& \quad \quad\quad player B: E\\ [1ex]
\hline
player A: F& \quad \quad\quad\quad (0,0,0)& \quad  \quad\quad\quad(1,-2,1)  \\ [1ex]
player A: E& \quad\quad \quad\quad(-2,1,1)& \quad \quad\quad\quad(1,1,-2) \\ [1ex]
\hline
\end{tabular}
\end{center}
\end{table}
\FloatBarrier

\FloatBarrier
\begin{table}[ht]
\begin{center}
\caption{Payoff matrix for the cooperative game, when player C plays strategy E.}\label{3}
\begin{tabular}{lll}
\hline
&\quad \quad\quad player B: F&\quad \quad\quad player B: E \\ [1ex]
\hline
player A: F&\quad \quad\quad\quad(1,1,-2)&\quad \quad\quad\quad(-2,1,1) \\ [1ex]
player A: E&\quad \quad\quad\quad(1,-2,1)&\quad \quad\quad\quad(0,0,0) \\ [1ex]
\hline
\end{tabular}
\end{center}
\end{table}
\FloatBarrier
\begin{flushleft}
where the left payoff is for player A, the middle is for player B, and the right is for player C.
\end{flushleft}

Let $\xi$ be a subset of $Q$. The players in $\xi$ form a harmonious coalition. The coalition appears as a single player with the same strategy. Suppose $\xi=\{B,C\}$, then $Q-\xi=\{A\},$ and $\Gamma_{\xi}$ is the coalition game given by the following payoff matrix:
\FloatBarrier
\begin{table}[ht]
  \centering
   \caption{Payoff matrix for the cooperative game, when players B and C form a coalition.}\label{4}
  \begin{tabular}{lll}
    \hline
&\quad \quad\quad Coalition $\xi$: FF&\quad \quad\quad Coalition $\xi$: EE\\ [1ex]
\hline
player A: F &\quad \quad\quad\quad \quad(0,0) &\quad \quad \quad\quad\quad(-2,2) \\ [1ex]
player A: E &\quad\quad \quad\quad\quad (-2,2) &\quad\quad \quad\quad\quad (0,0) \\ [1ex]
    \hline
  \end{tabular}
\end{table}
\FloatBarrier
\begin{flushleft}
where the left payoff is for player A, and the right payoff is for the coalition $\xi$.
\end{flushleft}

It is often useful and more realistic to expand the strategy space to include mixed strategies. Consider player A may play strategy F with probability $p$ and strategy E with probability $1-p$, and the coalition $\xi$ may play strategy F with probability $q$ and strategy E with probability $1-q$. Then the expected payoff of $Q-\xi$ is:
\begin{equation}\label{5}
  P_{Q-\xi}=4pq-2p-2q.
\end{equation}
At an equilibrium, the expected payoff can not depend on the probability $q$, then its partial derivative with respect to $q$ must be set to zero. This gives $p=\frac{1}{2}$, and the optimal mixed strategy for $Q-\xi$ becomes $\frac{1}{2} [F] + \frac{1}{2} [E]$. Similarly, for the coalition $\xi$ the optimal mixed strategy is $ \frac{1}{2} [FF] + \frac{1}{2} [EE]$.

The coalition $\xi$ is guaranteed to gain a payoff 1 for all strategies of the single player.
Then, the characteristic function of the coalition $\xi$ is $\upsilon(\Gamma_{\xi})= 1 $. Also, the characteristic function of the single player $Q-\xi$ is $\upsilon(\Gamma_{Q-\xi})=-1 $. Therefore, for any single player,
\begin{equation}\label{6}
  \upsilon(\{A\})=\upsilon(\{B\})=\upsilon(\{C\})= -1.
\end{equation}
And for any coalition,
\begin{equation}\label{7}
  \upsilon(\{A,B\})=\upsilon(\{B,C\})=\upsilon(\{C,A\})=1.
\end{equation}
Because the characteristic function of a coalition is the highest, players are motivated to form a coalition.
\section{Quantum Form}
We use Marinatto-Weber's scheme \cite{5} to quantise the symmetric cooperative game with harmonious coalition. A harmonious coalition with players B and C ia assumed. This means players B and C have to choose the same strategy. Therefore, we suppose an entangled initial quantum state as follow:
\begin{equation}\label{8}
 |\psi_{in}\rangle = C_{FFF}|FFF\rangle + C_{EFF}|EFF\rangle + C_{FEE}|FEE\rangle + C_{EEE}|EEE\rangle,
\end{equation}
where the coefficients $C_{FFF}, C_{EFF}, C_{FEE}, C_{EEE}\in \mathbb{C}$, such that
\begin{equation}\label{9}
  |C_{FFF}|^{2} +|C_{EFF}|^{2}+|C_{FEE}|^{2}+|C_{EEE}|^{2}=1.
\end{equation}
The ket $|FFF\rangle$ means player A plays strategy $|F\rangle$, player B plays strategy $|F\rangle$, and player C plays strategy $|F\rangle$, i.e. $|FFF\rangle=|F\rangle\otimes|F\rangle\otimes
|F\rangle$. Because of the harmony of the coalition the terms containing  $|EEF\rangle $, $|EFE\rangle$, $|FFE\rangle$ , and $|FEF\rangle$ are excluded from the initial state.

The unitary operators are defined as follows:
the identity operator $I=\left(
                       \begin{array}{cc}
                         1 & 0 \\
                         0 & 1 \\
                       \end{array}
                     \right)
$, and pauli-spin flip operator $\sigma=\left(
                                   \begin{array}{cc}
                                     0 & 1 \\
                                     1 & 0 \\
                                   \end{array}
                                 \right)
$, where
\begin{equation}\label{10}
\begin{split}
 & \sigma|F\rangle=|E\rangle, \quad \sigma|E\rangle=|F\rangle, \quad\sigma=\sigma^{\dagger}=\sigma^{-1},\\ &
                               I|F\rangle=|F\rangle,\quad I|E\rangle=|E\rangle,\quad I=I^{\dagger}=I^{-1}.
 \end{split}
  \end{equation}

After the initial state was introduced to players, the single player $\{A\}$ and the coalition $\{B,C\}$ carry out their strategies with probabilities $p$ and $q$ with the identity operator, and $1-p$ and $1-q$ with the pauli-spin flip operator, respectively on the initial state.

 The initial density matrix of $\psi_{in}$ is
 \begin{equation*}
   \rho_{in}=|\psi_{in} \rangle\langle\psi_{in}|.
 \end{equation*}

 \begin{equation}\label{11}\begin{split}
                             \rho_{in}=&[C_{FFF}|FFF\rangle + C_{EFF}|EFF\rangle + C_{FEE}|FEE\rangle + C_{EEE}|EEE\rangle] \\
                               & .[C^{*}_{FFF}\langle FFF| + C^{*}_{EFF}\langle EFF| + C^{*}_{FEE}\langle FEE| + C^{*}_{EEE}\langle EEE|].
                           \end{split}
 \end{equation}
 Then,
 \begin{equation}\label{12}
    \rho_{in}=\left(
                \begin{array}{cccccccc}
                  |C_{FFF}|^{2} & 0 & 0 & C_{FFF}C^{*}_{FEE} & C_{FFF}C^{*}_{EFF} & 0 & 0 & C_{FFF}C^{*}_{EEE} \\
                  0 & 0 & 0 & 0 & 0 & 0 & 0 & 0 \\
                  0 & 0 & 0 & 0 & 0 & 0 & 0 & 0 \\
                  C_{FEE}C^{*}_{FFF} & 0 & 0 & |C_{FEE}|^{2} & C_{FEE}C^{*}_{EFF} & 0 & 0 & C_{FEE}C^{*}_{EEE} \\
                  C_{EFF}C^{*}_{FFF}& 0 & 0 & C_{EFF}C^{*}_{FEE} & |C_{EFF}|^{2} & 0 & 0 & C_{EFF}C^{*}_{EEE} \\
                  0 & 0 & 0 & 0 & 0 & 0 & 0 & 0 \\
                  0 & 0 & 0 & 0 & 0 & 0 & 0 & 0 \\
                  C_{EEE} C^{*}_{FFF}& 0 & 0 & C_{EEE}C^{*}_{FEE} & C_{EEE}C^{*}_{EFF} & 0 & 0 & |C_{EEE}|^{2} \\
                \end{array}
              \right).
 \end{equation}
The final density matrix after players have played their strategies can be written as
\begin{equation}\label{13}
\begin{split}
                           \rho_{fin}= \quad&pq  \quad\quad\quad\quad\quad\quad\quad I_{A}\otimes I_{B}\otimes I_{C} \quad \rho_{in} \quad I^{\dag}_{A}\otimes I^{\dag}_{B}\otimes I^{\dag}_{C}  \\&
                         +p(1-q) \quad\quad\quad I_{A}\otimes \sigma_{B}\otimes \sigma_{C} \quad \rho_{in} \quad I^{\dag}_{A}\otimes \sigma^{\dag}_{B}\otimes \sigma^{\dag}_{C}  \\ &
                         + (1-p)q \quad\quad\quad \sigma_{A}\otimes I_{B}\otimes I_{C} \quad \rho_{in} \quad \sigma^{\dag}_{A}\otimes I^{\dag}_{B}\otimes I^{\dag}_{C} \\&
 + (1-p)(1-q) \quad \sigma_{A}\otimes \sigma_{B}\otimes \sigma_{C} \quad \rho_{in} \quad \sigma^{\dag}_{A}\otimes \sigma^{\dag}_{B}\otimes \sigma^{\dag}_{C}.  \end{split}
\end{equation}
The payoff operators for players A, B and C are
\begin{equation}\label{14}\begin{split}
                            P_{A,B,C}= \quad & a_{1},b_{1},c_{1}|FFF\rangle\langle FFF|+a_{2},b_{2},c_{2}|EFF\rangle\langle EFF| \\&
                             +a_{3},b_{3},c_{3}|FEE\rangle\langle FEE|+a_{4},b_{4},c_{4}|EEE\rangle\langle EEE|.
                          \end{split}
\end{equation}

where the constants $a_{i}, b_{i}$ and $c_{i}$, $i= 1, 2, 3, 4$ are the classical payoffs. The expected payoff functions of players A, B and C are
\begin{equation}\label{15}
\emph{\$}_{A,B,C}=Tr[P_{A,B,C} \quad \rho_{fin}].
\end{equation}
From the classical payoff of players A, B and C in Tables \ref{2} and \ref{3} the constants in Eq. (\ref{14}) become
\begin{equation}\label{16}\begin{split}
&a_1=a_4=0 ,\quad \quad \quad \quad\quad a_2=a_3=-2,\\&b_1=b_4=0 , \quad \quad \quad\quad\quad b_2=b_3=1,\\& c_1=c_4=0 ,\quad \quad \quad \quad\quad c_2=c_3=1.
\end{split}
\end{equation}
Then
\begin{equation}\label{17}
\begin{split}
\emph{\$}_{A}(p,q)=\quad&[|C_{FFF}|^{2}+|C_{EEE}|^{2}][4pq-2p-2q] \\&
+ [|C_{FEE}|^{2}+|C_{EFF}|^{2}][-4pq+2p+2q-2].
                          \end{split}
\end{equation}

\begin{equation}\label{18}
\begin{split}
  \emph{\$}_{B, C}(p,q)= \quad &[|C_{FFF}|^{2}+|C_{EEE}|^{2}][-2pq+p+q]\\&
  + [|C_{FEE}|^{2}+|C_{EFF}|^{2}][2pq-p-q+1].
  \end{split}
\end{equation}
 From Eqs. (\ref{17}) and (\ref{18}), it is obvious that $ \emph{\$}_{A}(p,q)+\emph{\$}_{B}(p,q)+\emph{\$}_{C}(p,q)=0$. This ensures the zero-sum property of the quantum game in our scheme.

\begin{figure}[ht]
  \begin{subfigure}[b]{0.4\textwidth}
    \includegraphics[width=\textwidth]{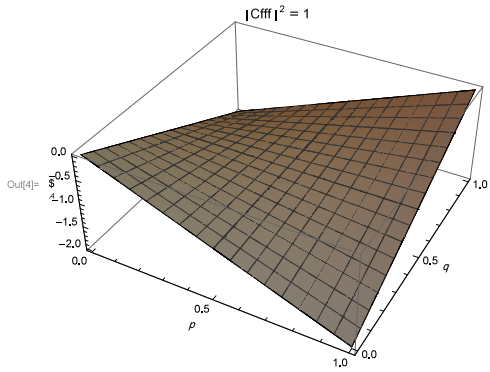}
    \caption{The expected payoff for player A.}
    \label{28}
  \end{subfigure}
  \hfill
  \begin{subfigure}[b]{0.4\textwidth}
    \includegraphics[width=\textwidth]{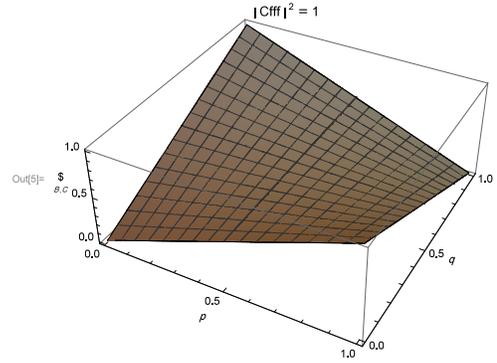}
    \caption{The expected payoff for players B and C.}
    \label{29}
  \end{subfigure}
  \caption{The expected payoffs of players when $|C_{FFF}|^{2}=1$.}
  \label{34}
\end{figure}

\begin{figure}[ht]
  \begin{subfigure}[b]{0.4\textwidth}
    \includegraphics[width=\textwidth]{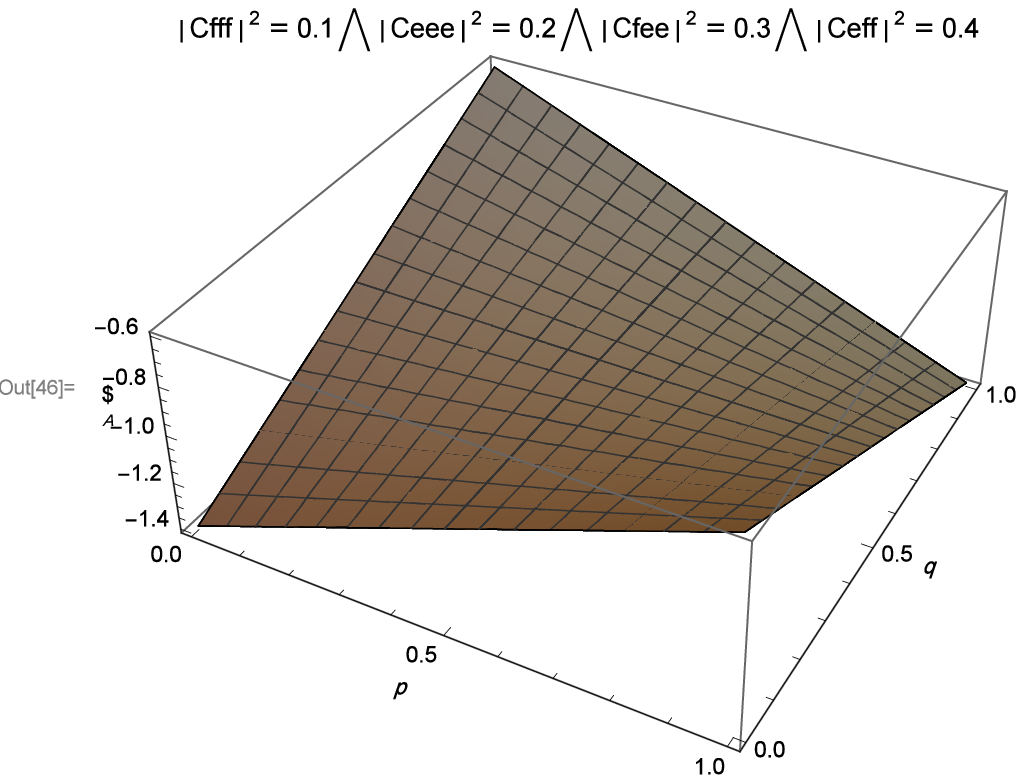}
    \caption{The expected payoff for player A.}
    \label{30}
  \end{subfigure}
  \hfill
  \begin{subfigure}[b]{0.4\textwidth}
    \includegraphics[width=\textwidth]{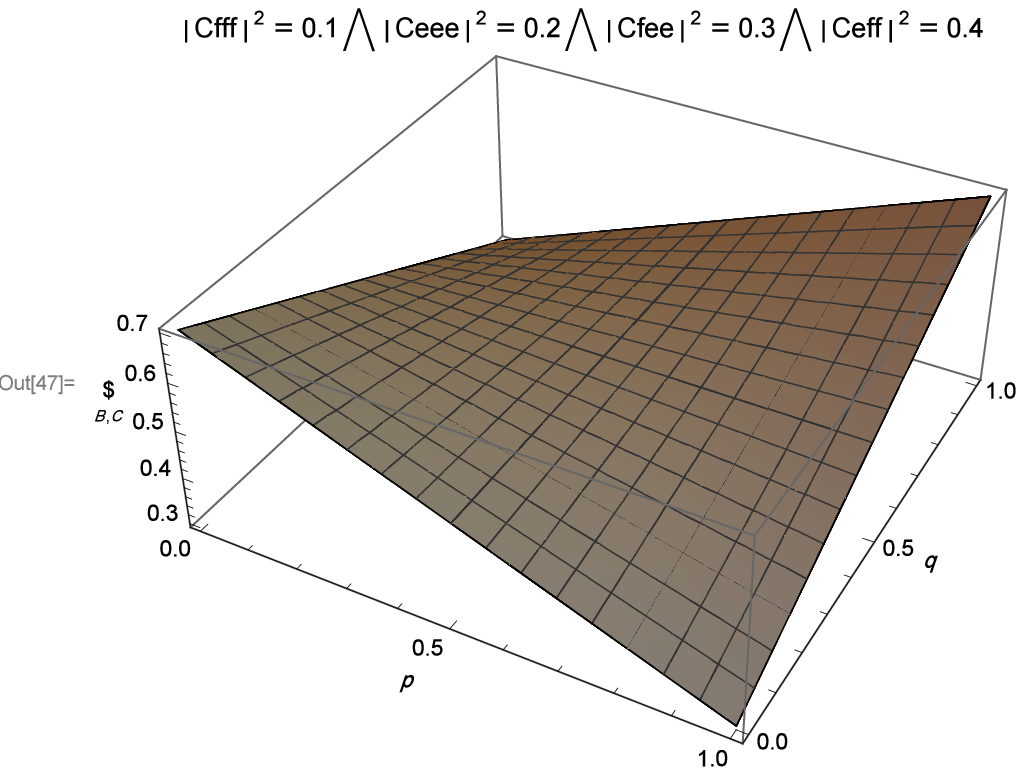}
    \caption{The expected payoff for players B and C.}
    \label{31}
  \end{subfigure}
  \caption{The expected payoffs of players when $|C_{FFF}|^{2}=0.1$, $|C_{EFF}|^{2}=0.4$, $|C_{FEE}|^{2}=0.3$ and $|C_{EEE}|^{2}=0.2$.}
  \label{35}
\end{figure}

\begin{figure}[ht]
  \begin{subfigure}[b]{0.4\textwidth}
    \includegraphics[width=\textwidth]{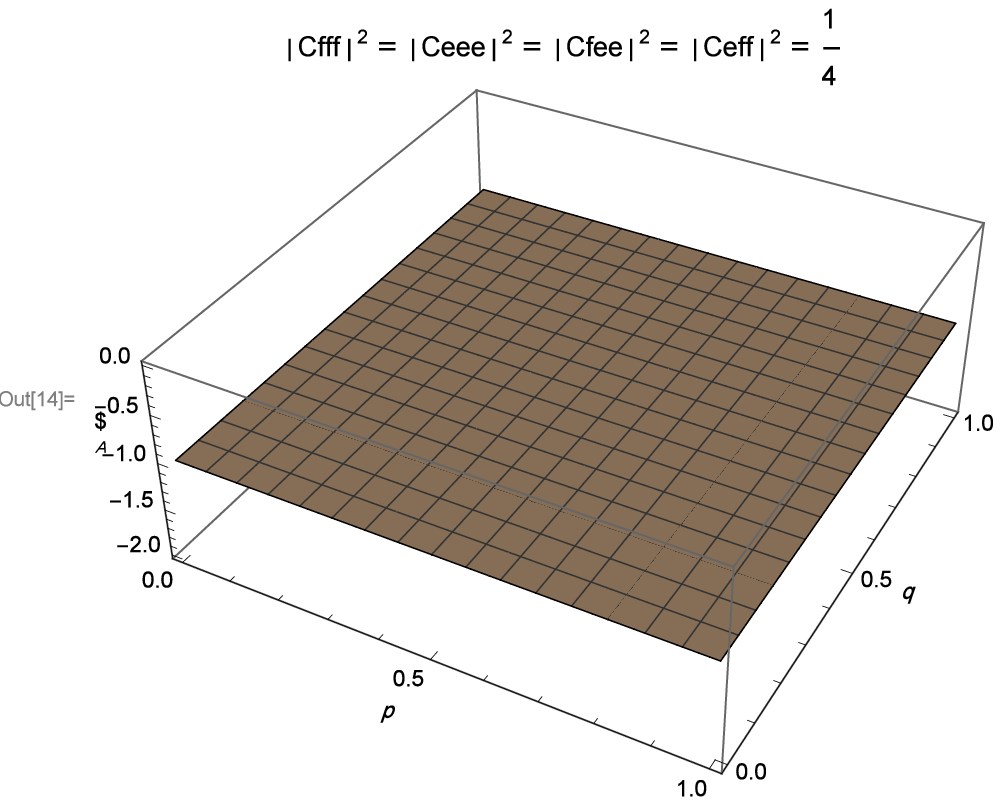}
    \caption{The expected payoff for player A.}
    \label{32}
  \end{subfigure}
  \hfill
  \begin{subfigure}[b]{0.4\textwidth}
    \includegraphics[width=\textwidth]{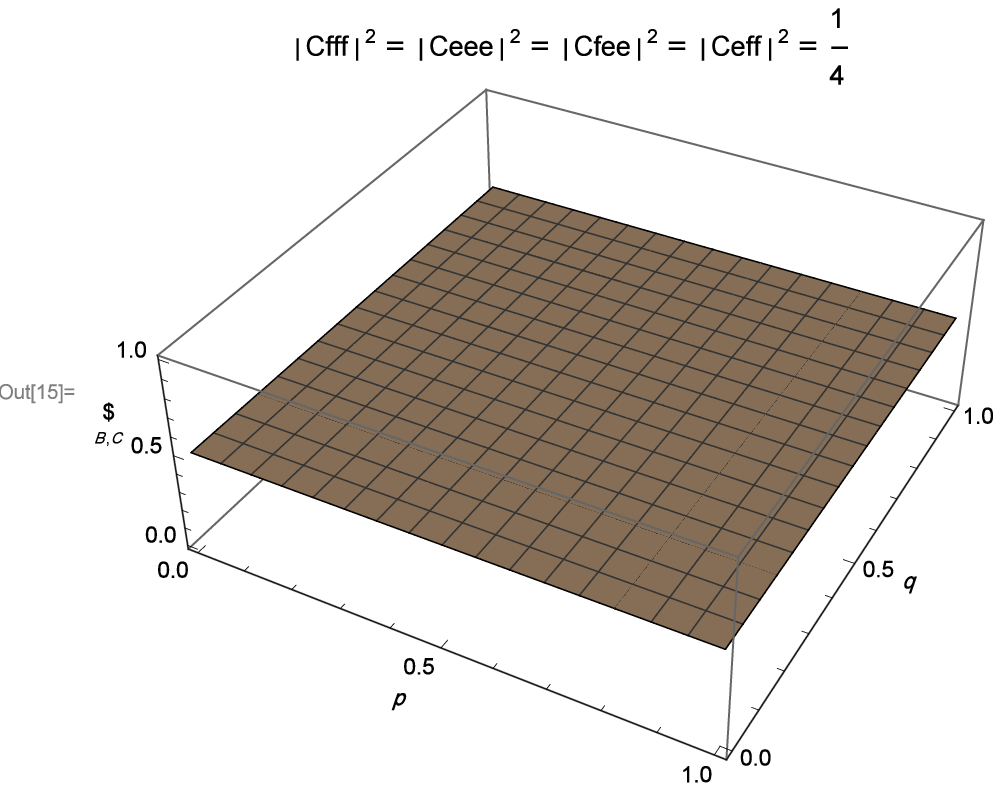}
    \caption{The expected payoff for players B and C.}
    \label{33}
  \end{subfigure}
  \caption{The expected payoffs of players when $|C_{FFF}|^{2}=|C_{EFF}|^{2}=|C_{FEE}|^{2}=|C_{EEE}|^{2}=1/4$.}
  \label{36}
\end{figure}

To investigate how the quantum aspect affects the outcome of the game, the expected payoffs $ \emph{\$}_{A}$ and $\emph{\$}_{B,C}$ are plotted with respect to both the probabilities $p$ and $q$ at different values for the constants $|C_{FFF}|^{2}$, $|C_{EFF}|^{2}$, $|C_{FEE}|^{2}$, and $|C_{EEE}|^{2}$ that corresponding to different initial states. In Fig. \ref{34}, $|C_{FFF}|^{2}=1$, the initial state is non entangled. The classical game payoffs appear as limiting points of the payoffs of the quantum version for non entangled initial states. Fig. \ref{35} shows the relation between both $\emph{\$}_{A}$ and $\emph{\$}_{B,C}$ and the probabilities $p$ and $q$ for an entangled initial state. The payoffs for the maximally entangled initial state are shown in Fig. \ref{36}. Because of the zero-sum property, the behavior of $\emph{\$}_{B}+\emph{\$}_{C}$ is the inverse of the behavior of $\emph{\$}_{A}$ in all cases. Figs. \ref{34}-\ref{36} clearly show the dependence of the expected payoff in the quantum version on the initial quantum state and the players strategies.

Now, let the coalition $\xi$ play the mixed strategy $s[FF]+(1-s)[EE]$, and the single player $Q-\xi$ play the mixed strategy $t[F]+(1-t)[E]$.
In this case the payoff of the coalition $\xi$ is obtained as
\begin{equation}\label{19} \begin{split}
                             \textbf{ P}_{\xi}=&stP_{\xi}[FFF]+s(1-t)P_{\xi}[EFF] \\ &
                                +(1-s)tP_{\xi}[FEE]+(1-s)(1-t)P_{\xi}[EEE],
                           \end{split}
\end{equation}
where $P_{\xi}[FFF]$ is the payoff of $\xi$ when all players play the strategy $[F]$, i.e. $p=1$ and $q=1$.
Now from Eq. (\ref{18}), we get
\begin{equation}\label{20}
  P_{\xi}[FFF]=P_{B}[11]+P_{C}[11]=2[|C_{FEE}|^{2}+|C_{EFF}|^{2}]=P_{\xi}[EEE]  ,
\end{equation}

\begin{equation}\label{21}
  P_{\xi}[EFF]=P_{B}[01]+P_{C}[01]=2[|C_{FFF}|^{2}+|C_{EEE}|^{2}]=P_{\xi}[FEE]  .
\end{equation}
Therefore, from Eq. (\ref{19}), we get
\begin{equation}\label{22} \begin{split}
                          \textbf{ P}_{\xi}=&2st[|C_{FEE}|^{2}+|C_{EFF}|^{2}]+2s(1-t)[|C_{FFF}|^{2}+|C_{EEE}|^{2}]     \\&
+2(1-s)t[|C_{FFF}|^{2}+|C_{EEE}|^{2}]+2(1-s)(1-t)[|C_{FEE}|^{2}+|C_{EFF}|^{2}] .
                           \end{split}
\end{equation}

 To get the optimal payoff, the partial derivative with respect to $t$ must be set to zero. It gives $s=\frac{1}{2}$. Similarly, for the single player $t=\frac{1}{2}$. Therefore, the characteristic function of the coalition $\xi$ becomes
\begin{equation}\label{23}
\begin{split}
     \upsilon(\Gamma_{\xi})&=\upsilon(\{B, C\}) \\&
     =|C_{FFF}|^{2} +|C_{EFF}|^{2}+|C_{FEE}|^{2}+|C_{EEE}|^{2}=1.
   \end{split}
\end{equation}
Similarly, the payoff of the single player $Q-\xi$ is
\begin{equation}\label{24} \begin{split}
                             \textbf{P}_{Q-\xi}=&stP_{Q-\xi}[FFF]+s(1-t)P_{Q-\xi}[EFF]  \\&
                                +(1-s)tP_{Q-\xi}[FEE]+(1-s)(1-t)P_{Q-\xi}[EEE].
                           \end{split}
\end{equation}
From Eq. (\ref{17}), we get
\begin{equation}\label{25}
   P_{Q-\xi}[FFF]=P_{A}[11]=-2[|C_{FEE}|^{2}+|C_{EFF}|^{2}]=P_{Q-\xi}[EEE] ,
\end{equation}

\begin{equation}\label{26}
   P_{Q-\xi}[EFF]=P_{A}[01]=-2[|C_{FFF}|^{2}+|C_{EEE}|^{2}]=P_{Q-\xi}[FEE] .
\end{equation}
Then
\begin{equation}\label{27}
  \upsilon(\Gamma_{Q-\xi})=\upsilon(\{A\})=-[|C_{FFF}|^{2} +|C_{EFF}|^{2}+|C_{FEE}|^{2}+|C_{EEE}|^{2}]=-1
\end{equation}

From Eqs. (\ref{23}) and (\ref{27}), the quantised game is a zero-sum game, and the players have the advantage to form a coalition.

 A harmonious coalition with players A and B can be assumed using the following initial state:
 \begin{equation}\label{37}
  |\psi_{in}\rangle = C_{FFF}|FFF\rangle + C_{FFE}|FFE\rangle + C_{EEF}|EEF\rangle + C_{EEE}|EEE\rangle .
\end{equation}
 Using the same procedures, we get:
\begin{equation}\label{38}
 \upsilon(\{A, B\})=1 \quad and \quad \upsilon(\{C\})=-1.
\end{equation}

 Similarly, for a coalition with players A and C, we have
\begin{equation}\label{39}
  \upsilon(\{A, C\})=1 \quad and \quad \upsilon(\{B\})=-1.
\end{equation}
These findings conserve the results of the classical form.
But these results are in disagreement with the results of Iqbal and Toor \cite{6}, that both the motivation for forming coalition and the zero-sum property are lost in the quantised version.

The differences between Iqbal and Toor's scheme \cite{6} and ours are summarized in the following. We assume an initial state that allows to form a coalition beforehand. Introducing the concept of harmonious coalition that makes the coalition behaves as a single player.

\section{Conclusion}
Both classical and quantum versions of the cooperative game may have similar features; but they differ in details. The classical version depends only on a set of few pure classical strategies. In quantum version, the strategy space is extended to a Hilbert space of strategies. The initial quantum state prepared by an arbiter plays a crucial role in the quantum form. For some initial states, cooperation and forming a coalition are very important (our work). For some other initial states, cooperation is useless \cite{6}.

 The payoff in the quantum version depends on both the initial state and the players strategies (see Figs. \ref{34}-\ref{36}). The classical version of the game appears as a limiting point of the quantum version (see Fig. \ref{34}).
The initial quantum state acts like a social contract that organizes the relations between quantum players.

\section*{Acknowledgements:}
We would like to thank the referees for their helpful comments.

\end{document}